\documentclass[a4paper,11pt]{article}
\pdfoutput=1 
\usepackage{jheppub}         
\usepackage{amssymb,amsfonts,amscd,latexsym,amsbsy, amsmath,bm}
\usepackage[caption = false]{subfig}
\usepackage[section]{placeins} 
\usepackage{multirow, booktabs, amsthm, xfrac, fix-cm}
\usepackage{url, amsmath, amssymb, color, setspace, tabu} 
\usepackage[T1]{fontenc} 
\usepackage{etoolbox}
\numberwithin{equation}{section}
\usepackage{dsfont}
\usepackage[usenames,dvipsnames]{xcolor}
\usepackage{slashed}
\usepackage{verbatim} 
\usepackage{graphicx} 
\usepackage{caption}
\usepackage{wrapfig} 
\usepackage{anyfontsize}
\usepackage{calligra}
\usepackage{cancel}
\graphicspath{{./graficas/}} 
\DeclareGraphicsExtensions{.jpg,.eps,.jpeg,.pdf,.png,.2} 
\usepackage{float}
\usepackage[utf8]{inputenc}
\usepackage{stackrel}
\usepackage{mathrsfs}
\usepackage[utf8]{inputenc}
\usepackage{float}%
\usepackage{tikz}
\usetikzlibrary{trees}
\usetikzlibrary{arrows,positioning,decorations.pathreplacing,shapes,snakes,decorations.pathmorphing,decorations.markings}
\tikzset{>=latex}
\usepackage{longtable}
\usepackage{epic}
\usepackage{eepic}
\usepackage{array}
\usepackage{pgf,pgfarrows}
\usepackage{tikz}
\usepackage{multirow}
\usepackage{float}
\usetikzlibrary{shapes}
\usetikzlibrary{plotmarks}
\usepackage{wrapfig}
\usepackage{enumerate}
\usepackage[curve]{xy}
\usepackage{stmaryrd}
\usepackage{mathtools}
\usepackage{float}
\usepackage{wrapfig}

\newcommand{\Bigpsi}[3]{\phantom{\Psi}_2 \kern -.05em
\Psi_2\left(\genfrac{}{}{0pt}{}{#1}{#2}\biggl|#3\right)}
\newcommand{\bea}{\begin{eqnarray}}
\newcommand{\eea}{\end{eqnarray}}
\newcommand{\beq}{\begin{equation}}
\newcommand{\eeq}{\end{equation}}

\renewcommand{\textcolor}[1]{}

\usepackage{mathtext}
\usepackage{stackrel}
\usepackage[utf8]{inputenc}
          \AfterPreamble{
           \hypersetup    {
       linkcolor=blue,
        urlcolor=blue,
        citecolor=blue,
  pdfauthor={Shahriyar Jafarzade, Zainab Nazari},
   }}

\title{\boldmath {A New Integrable Ising-type Model \protect \\from 2d $\mathcal{N}$=(2,2) Dualities} }
\author[\Yup]{Shahriyar Jafarzade}
\author[\vartriangle]{and Zainab Nazari}
\affiliation[\Yup]{Department of Physics, Mimar Sinan Fine Arts University, Bomonti, 34380, Istanbul, Turkey}
\affiliation[\Yup]{Department of Mathematics, Khazar University, AZ1096, Baku, Azerbaijan}
\affiliation[\Yup]{Institute of Radiation Problems ANAS, AZ1143, Baku, Azerbaijan}
\affiliation[\vartriangle]{Department of Physics, Bo\u{g}azi\c{c}i University, Bebek, 34342, Istanbul, Turkey}
\affiliation[\vartriangle]{Feza G\"{u}rsey Center for Physics and Mathematics, Kandilli, 34684, Istanbul, Turkey}

\emailAdd{shahriyar.jzade(at)gmail.com}
\emailAdd{zainab.nazari(at)boun.edu.tr}

\abstract{We show that the equality of 2d $\mathcal{N}$=(2,2) supersymmetric indices in Seiberg-type duality leads to a new integrable Ising-type model. The emergence of the new model is the result of  correspondence between the supersymmetric $SU(2)$ quiver gauge  theories and the Yang-Baxter equation. Using this correspondence, we solve the star-triangle relation and obtain the two-dimensional exactly solvable spin model. The model corresponding to our solution possesses continuous spin variables on the circle and the Boltzmann weights are demonstrated in terms of the Jacobi theta function. Using the solution of the star-triangle relation, we also construct  interaction-round-a-face and  vertex  models.}

\begin{document} 

\maketitle

\flushbottom

\section{Introduction}
In the past few years, some remarkable connections between  supersymmetric
 gauge theories and integrable statistical models have been observed \cite{Spiridonov:2010em,Yamazaki:2012cp,Yagi:2015lha,Nieri:2013vba,Costello:2013zra,Witten:2016spx,Nekrasov:2009ui,Orlando:2010uu}. One of these connections represents the correspondence  between the quiver gauge theories and the integrable lattice models, where supersymmetric duality on the gauge theory side can be interpreted as the integrability on the lattice side \cite{Spiridonov:2010em, Kels:2015bda,Yagi:2015lha,Gahramanov:2016ilb,Gahramanov:2015cva, Yamazaki:2015voa, Yamazaki:2013nra,Maruyoshi:2016caf,Yagi:2017hmj}. 
 
 The correspondence between the supersymmetric  gauge theories and  the  Yang-Baxter equation, known as the gauge/YBE correspondence, has shed light on exactly solvable models \cite{Yamazaki:2013nra}. Exactly solvable models play a major role in the investigation of critical phenomena in statistical mechanics \cite{baxter2007exactly}, where the YBE as a highly over-constrained equation is the key structural element leading to the exact solvability.
 According to this  correspondence, partition functions or supersymmetric indices of supersymmetric gauge theories are identified with the partition functions of two-dimensional exactly solvable models in statistical mechanics. In other words, quiver diagrams from the gauge theory point of view appear as the lattice on which the spin  models are represented \cite{Yamazaki:2013nra}. The manifestation of this correspondence, can also be understood in another way through topological quantum field theory \cite{Yagi:2015lha}.  All recent models which have been discovered using this correspondence, are listed in \cite{Jafarzade-2017}.
 In this paper, we present a new integrable Ising-type model via the identification of the 2d $\mathcal{N}$=(2,2) supersymmetric indices of the Seiberg-type duality with the star-triangle relation. The Seiberg-type duality corresponding to the new model is between the $SU(2)$ gauge theory with 6 flavors, and its dual theory with 15 mesons. The star-triangle relation as the simplest form of the YBE reads
  \begin{align}\label{Strr}
     \int d \sigma \mathcal{S}( \sigma ) \mathcal{W}_{\eta-\alpha}(\sigma _{i},\sigma)& \mathcal{W}_{\eta - \beta}( \sigma _{j}, \sigma) \mathcal{W}_{ \eta - \gamma}( \sigma , \sigma_{k})=\\\nonumber
    & \qquad \mathcal{R}(\alpha , \beta,\gamma)\mathcal{W}_{ \alpha }( \sigma _{j}, \sigma_{k})\mathcal{W}_{ \beta}( \sigma _{i}, \sigma_{k})\mathcal{W}_{ \gamma}( \sigma _{j}, \sigma_{i}),
\end{align}
where $\mathcal{W}$'s are the Boltzmann weights, $\mathcal{S}$ stands for the self-interactions and $\mathcal{R}$ is called the R-factor of the two-dimensional integrable model \cite{Bazhanov:2016ajm,Baxter:1997tn}.

There are three types of the YBE in the context of statistical mechanics, the star-triangle relation, the interaction-round-a-face (IRF) and the vertex types. In this letter, we present solutions to all three types. However, our main focus is on the star-triangle solution, since the other types of the YBE can be obtained from it, and the solution to the star-triangle relation is a sufficient condition to get an integrable Ising-type model. Ising-type models are distinguished according to their spin variables and the types of Boltzmann weights. Boltzmann weights determine the interactions between the spins, where the spins  are spaced in the edges of two-dimensional square lattice.

The new solution to the star-triangle  relation\footnote{One may need to check whether it is a limit of the solution presented by Kels in \cite{Kels:2015bda}, which contains most of the previously known solutions as a special limit.} in terms of the Jacobi theta function involves continuous spin variables, varying in the range $0\leq \sigma< 2\pi$, and is as follows
 
 \begin{equation}\label{IB-W}\nonumber
     \,\mathcal{W}_{\alpha}(\sigma_i,\sigma_j)=\frac{\theta( e^{-\alpha-\eta\mp i(\sigma_i\pm \sigma_j)};q)}{\theta(e^{\alpha-\eta\pm i(\sigma_i\pm \sigma_j)};q)},\qquad\qquad\qquad\,
\end{equation}
\begin{equation}
     \mathcal{S}(\sigma)=\frac{1}{4\pi}\Big(\frac{(q;q)^2_{\infty}}{\theta(y;q)}\Big)  \frac{\theta(e^{\pm 2i\sigma};q)}{\theta(e^{-2\eta\pm 2i\sigma};q)},\, 
\end{equation}
\begin{equation}\nonumber
  \qquad \qquad \mathcal{ R}(\alpha,\gamma,\beta)= \frac{\theta(e^{-2\alpha-2\gamma};q)}{\theta(e^{-2\beta};q)} \frac{\theta(e^{-2\beta-2\gamma};q)}{\theta(e^{-2\alpha};q)} \frac{\theta(e^{-2\beta-2\alpha};q)}{\theta(e^{-2\gamma};q)},\,\,\,
\end{equation}
where the Jacobi theta function is defined as
\begin{align} \label{theta}
    \theta(z;q):= \prod_{i=0}^{\infty}(1-zq^i)(1-z^{-1}q^{i+1}),
\end{align}
and  q-Pochhammer is 
\begin{align}\label{q-Poch}
  (z;q)_\infty:=\prod_{i=0}^{\infty}(1-zq^i).\qquad \qquad\quad\,\,\,\,
\end{align}
 According to the gauge/YBE correspondence, the interpretation of the star-triangle relation  elements (\ref{Strr}), in terms of supersymmetric theories are as follows. $S(\sigma)$ refers to the contribution of the vector multiplets, $\mathcal{W}_\alpha(\sigma_i,\sigma_j)$ refers to the contribution of the chiral multiplets, $\sigma$ is fugacity of the gauge group. Spectral parameters $\alpha, \beta, \gamma$ are related to the R-charges and $\sigma_{i,j,k}$ are fugacities of the flavor group \cite{Yamazaki:2012cp,Terashima:2012cx}.

To elaborate on the  introduction, we firstly go through the gauge theory discussions and then look for the statistical mechanical resemblance with the first part, and in the end present the new solutions which we achieve using the gauge/YBE correspondence. Therefore, the rest of the paper is organized as follows. In Section \ref{2d sup}, supersymmetric indices and Seiberg-type duality are briefly reviewed. In Section \ref{Integ}, integrability and the star-triangle relation are studied. In Section \ref{sol}, the new Ising-type model and its high temperature limit is presented, afterwards the IRF and the vertex type models and the properties of the Boltzmann weights are demonstrated.  The structure of the paper ends with Section \ref{fu}, which contains possible future directions. In Appendix \ref{appa}, some useful properties of the Jacobi theta function are given. In Appendix \ref{proof}, we present a proof for the 2d Seiberg-type duality. In Appendix \ref{AIRF}, the detailed calculation of IRF model is presented.

\section{2d $\mathcal{N}$=(2,2) Dualities}\label{2d sup}

The identification of supersymmetric indices in Seiberg-type duality with the star-triangle form of the YBE leads to a new Ising-type model. In this section, we discuss the supersymmetric index technique which is considered as the most convenient way to check the existence of Seiberg-type dualities \cite{Spiridonov:2008zr}. The first part is devoted to introduce the supersymmetric indices of 2d $\mathcal{N}$=(2,2) theories, and in the second part, we summarize the main ideas behind Seiberg-type duality.
\subsection{ 2d Supersymmetric Index} 

The 2d supersymmetric index is a generalization of the Witten index, defined on $S^{1} \times S^{1}$, and it is a non-trivial function of flavor and supersymmetric fugacities \cite{Nakayama:2011pa,Gadde:2013wq,Benini:2013nda, Gadde:2013ftv}. 
The Witten index is an effective tool to check whether the supersymmetry is spontaneously broken or not \cite{Witten:1982df}, and it represents the difference between the number of bosonic and fermionic ground states. \\For simplicity, let us start with the  index of  1d supersymmetric quantum mechanics 
\begin{equation} \label{W-Index}
\mathcal{I} =\text{Tr}_\mathbb{H}\Big[(-1)^{F}e^{- \beta \big\{Q, Q^\dag\big\}} \Big],
\end{equation}
 where $Q$, $Q^{\dagger}$ are the supercharges with the Hamiltonian, $\mathcal{H}=\{Q, Q^{\dagger}\}$. $(-1)^{F} $ is the fermionic number operator. The trace is taken over all states which survive with the cancellation of bosonic and fermionic states, so the trace localizes to the part with Hamiltonian $ \mathcal{H}=0$, subset of the Hilbert space $\mathbb{H}$.\\
Now, let us define the supersymmetric index of the 2d $\mathcal N=(2,2)$ gauge theories as
\begin{equation}\label{Ind}
\mathcal{I}( \{a_j\},q,y )=\text{Tr}_\mathbb{H}\Big[ (-1)^F e^{-\beta(2H_r+J_r)}q^{H_l}y^{J_l} \prod_j a_j^{F_j} \Big],
\end{equation}
where $J_r$ ($J_l$) and $H_r$ ($H_l$) are the right(left)-moving R symmetry and right(left)-moving conformal dimension, respectively. In this case $Q^{\pm}_{l,r}$ are the generators of the supersymmetric algebra. $F_j$ are the generators of global symmetries which commute with $Q^{\pm}_{l,r}$, and $a_j$ are the corresponding fugacities. Likewise the Witten index (\ref{W-Index}), the trace  localizes to the $(2H_{r}+J_{r}=0)$, subset of the Hilbert space $\mathbb{H}$. The structure of the 2d $\mathcal N=(2,2)$ supersymmetric index has  composed of gauge vector and chiral matter multiplets\footnote{Though, there are also non-zero contributions from the Chern-Simons and the Fayet-Iliopoulos classical action, we ignore them. The reason is that, the Seiberg-type dual theories in the context of gauge/YBE correspondence are without these contributions.}, and has the following form
\begin{equation}\label{ind}
  \mathcal{I}(\{a_j\},q,y)= \frac{1}{|W|}\int \prod_{i=1}^{\text{rankG}}\frac{dz_i}{2\pi iz_i}  \mathcal{Z}_{V,G}(z_i,q,y) \prod_{\Phi} \mathcal{Z}_{\Phi,\rho}(z_i,a_j,q,y),
\end{equation}
where the prefactor $|W|$ is the order of the Weyl group of gauge group $G$.\\ The contribution of the chiral multiplet with vector-like R-charge $r$, which transforms
in a representation $\mathfrak{R}$ of the gauge and flavor group G is
\begin{align}\label{Chiral}
  \mathcal{Z}_{\Phi,\rho}=\prod_{\rho\in \mathfrak{R} }\frac{\theta(y^{r/2-1}a^{\rho};q) }{\theta(y^{r/2}a^{\rho};q)},\qquad \qquad\qquad\quad
\end{align}
where the product is over the weights $\rho$ of the representation $\mathfrak{R}$. $a^{\rho}:=e^{2\pi i \rho(u)}$, where $u$ is the element of Cartan subalgebra of the flavour group.\\ The contribution from the vector multiplet  with gauge group $G$ reads
\begin{align}\label{Vec}
  \mathcal{Z}_{V,G}=\Big(\frac{(q;q)^2_{\infty}}{\theta(q^{\frac12}y^{-1};q)}\Big)^{\text{rankG}}\prod_{\alpha \in G } \frac{\theta(a^{\alpha};q)}{\theta(y^{-1}a^{\alpha};q)},
\end{align}
where $a^{\alpha}:= e^{2\pi i \alpha(z)}$, and the product is over the roots $\alpha$ of the gauge group  $G$\footnote{Here, we define the index in the NSNS sector \cite{Gadde:2013ftv}, it can also be defined in the RR sector as in \cite{Benini:2013nda,Benini:2013xpa}.}.

\subsection{Seiberg-type Duality} \label{duality}

  Seiberg-type duality plays the main role in the discussion of gauge theory side in gauge/YBE correspondence. Therefore, here we summarize some of its main features.\\According to the original Seiberg  duality, 4d $\mathcal{N}=1$ supersymmetric gauge theories with different ultraviolet behavior flow to the same infrared fixed point, where the dual theories in the low energy  describe the same physics.  The UV dual theories are, an \emph{electric} theory with an $SU(N_c$) gauge group and $N_f$ flavors, and a \emph{magnetic} theory with an $SU(N_f-N_c)$ gauge group with $N_f$ flavors and an additional gauge invariant massless field. There are several features of this duality.  For instance, the quarks and gluons of one theory corresponds to the solitons of the elementary fields of the other one.
The weak coupling region of the electric side can be interpreted as the strong coupling region of the magnetic side, etc.  The simplest prediction of the duality is that the moduli spaces of supersymmetric vacua of the dual theories are the same \cite{Seiberg:1994pq,Hori:2006dk,Yamazaki:2013fva}.

 There are  many  generalizations of this duality in the literature with complicated matter contents, different gauge and flavor groups in various dimensions. One generalization applies to quiver gauge theories in which the flavor symmetries are also gauged \cite{Berenstein:2002fi}. Supersymmetric indices in so-called Seiberg-type dualities can provide non-trivial integral identities which can be used as a mathematical tool to check the duality of non-abelian
gauge theories as in \cite{Dolan:2008qi,Spiridonov:2009za,Dolan:2011rp,Spiridonov:2011hf,Gahramanov:2013rda,Gahramanov:2014ona,Gahramanov:2015tta,Gahramanov:2016wxi}.
The Seiberg-type duality that we consider in this paper which corresponds to the star-triangle relation also has two sides, the electric-like and the magnetic-like sides. The electric theory part  has gauge group $SU(2)$, with 6 flavors (or alternatively, 3 fundamental and 3 anti-fundamental flavors). In the magnetic  theory part, there is no gauge symmetry and there are 15  chiral multiplets (mesons) which  transform under totally antisymmetric tensor representation of the flavor group. It can be seen that the chiral multiplets transform under fundamental representations of the gauge and flavor group, and the vector multiplets transform in the adjoint representation of the gauge group.

For the purpose of this letter,  a version
in which its  matter contents are essentially the same as the one in 4d $\mathcal{N}$=1 Seiberg duality is considered.
The Seiberg-type duality of 2d $\mathcal{N}$=(2,2) supersymmetric theory that we consider here, is a gauged linear sigma model, and its Lagrangian can
be obtained from the dimensional reduction of 4d $\mathcal{N}$ = 1 theory, as discussed in \cite{Honda:2015yha}. It can be shown that the invariance of supersymmetric indices in the dual theories are equivalent to the integrability of the spin models, e.g. \cite{Yamazaki:2013nra}. This equality leads to the construction of the integrable systems.

\section{Integrability } \label{Integ}

In this section, the lattice spin model is described and its partition function with continuous spin variables is given. The star-triangle relation in terms of rapidity variables and spectral parameters are presented, and some properties of the Boltzman weights for the new integrable model are represented.

\subsection{Lattice Spin Model }
\begin{figure}[H]
\centering
\begin{tikzpicture}[scale=0.5,thick]
     \draw [->,dashed,color=black] (-4,-4) node[left,scale=.8pt]  {$p_1$}    to (2.5,-4) ;
     \draw [->,dashed,color=black] (-4,-3) node[left,scale=.8pt]  {$p_2$}   to (2.5,-3) ;
     \draw [->,dashed,color=black] (-4,-2) node[left,scale=.8pt]  {$p_3$}   to (2.5,-2) ;
     \draw [->,dashed,color=black] (-4,-1) node[left,scale=.8pt]  {$p_4$}   to (2.5,-1) ;
     \draw [->,dashed,color=black] (-4,0) node[left,scale=.8pt]  {$p_5$}  to (2.5,0) ;
          \draw [->,dashed,color=black] (-3,-5) node[below,scale=.8pt]  {$q_1$}   to (-3,1.5)   ;
      \draw [->,dashed,color=black] (-2,-5) node[below,scale=.8pt]  {$q_2$}  to (-2,1.5) ; 
      \draw [->,dashed,color=black] (-1,-5) node[below,scale=.8pt]  {$q_3$}  to (-1,1.5) ;
        \draw [->,dashed,color=black] (-0,-5) node[below,scale=.8pt]  {$q_4$}  to (0,1.5) ;
         \draw [->,dashed,color=black] (1,-5)node[below,scale=.8pt]  {$q_5$}   to (1,1.5) ;

                       \node (h) at (-3.5,-4.5) [circle,draw=blue!100, fill=blue!100] [scale=0.4pt] {};
     \node (h) at (-3.5,-2.5) [circle,draw=blue!100, fill=blue!100] [scale=0.4pt] {};
     \node (h) at (-3.5,-0.5) [circle,draw=blue!100, fill=blue!100] [scale=0.4pt] {};
              \node (h) at (-2.5,-3.5) [circle,draw=blue!100, fill=blue!100] [scale=0.4pt] {};
          \node (h) at (-2.5,0.5) [circle,draw=blue!100, fill=blue!100] [scale=0.4pt] {};
          \node (h) at (-1.5,-4.5) [circle,draw=blue!100, fill=blue!100] [scale=0.4pt] {};
          \node (h) at (-1.5,-2.5) [circle,draw=blue!100, fill=blue!100] [scale=0.4pt] {};
          \node (h) at (-1.5,-0.5) [circle,draw=blue!100, fill=blue!100] [scale=0.4pt] {};
          
          \node (h) at (-0.5,-3.5) [circle,draw=blue!100, fill=blue!100] [scale=0.4pt] {};
          \node (h) at (-0.5,-1.5) [circle,draw=blue!100, fill=blue!100] [scale=0.4pt] {};
                    \node (h) at (-0.5,0.5) [circle,draw=blue!100, fill=blue!100] [scale=0.4pt] {};
        \node (h) at (0.5,-4.5) [circle,draw=blue!100, fill=blue!100] [scale=0.4pt] {};
         \node (h) at (0.5,-2.5) [circle,draw=blue!100, fill=blue!100] [scale=0.4pt] {};
        \node (h) at (0.5,-0.5) [circle,draw=blue!100, fill=blue!100] [scale=0.4pt] {};
                   \node (h) at (1.5,-3.5) [circle,draw=blue!100, fill=blue!100] [scale=0.4pt] {};
           \node (h) at (1.5,-1.5) [circle,draw=blue!100, fill=blue!100] [scale=0.4pt] {};
            \node (h) at (1.5,0.5) [circle,draw=blue!100, fill=blue!100] [scale=0.4pt] {};
                                \node (h) at (-2.5,-1.5) [circle,draw=blue!100, fill=blue!100] [scale=0.4pt] {};

		\draw [thin,color=cyan] (-3.5,-4.5) to (1.5,0.5) ;           				
		\draw [thin,color=cyan] (-1.5,-4.5) to (1.5,-1.5) ;                 					
		\draw [thin,color=cyan] (0.5,-4.5) to (1.5,-3.5) ;   													
		\draw [thin,color=cyan] (-3.5,-2.5) to (-0.5,0.5) ;	
		\draw [thin,color=black] (-3.5,-2.5) to (-1.5,-4.5) ;		
		\draw [thin,color=black] (-3.5,-0.5) to (0.5,-4.5) ;
	\draw [thin,color=cyan] (-3.5,-0.5) to (-2.5,.5) ;
\draw [thin,color=black] (-0.5,0.5) to (1.5,-1.5) ;	
\draw [thin,color=black] (1.5,-3.5) to (-2.5,0.5) ;

	\end{tikzpicture}
		
\caption{Lattice spin model}
\label{lspin}
 \end{figure}
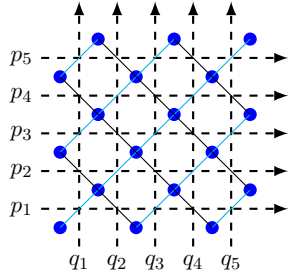

\begin{figure}[H]
 \centering
\begin{tikzpicture}[scale=0.5,thick]
             \draw [thin,color=black] (-1.5,0) node[left,scale=1.pt] {$\sigma_i$} to (1.5,0) node[right,scale=1.pt] {$\sigma_j$};
   \node (h) at (1.5,0) [circle,draw=blue!100, fill=blue!100] [scale=0.4pt] {};
   \node (h) at (-1.5,0) [circle,draw=blue!100, fill=blue!100] [scale=0.4pt] {};
          \draw [->,dashed,color=black] (-1.5,-1.5) node[below,scale=.8pt] {$p_m$} to (1.5,1.5) ;
       \draw [->,dashed,color=black] (1.5,-1.5) node[below,scale=.8pt]{$q_n$} to (-1.5,1.5) ;      
		\end{tikzpicture}
		\qquad \qquad
\begin{tikzpicture}[scale=0.5,thick]
\draw [thin,color=cyan] (0,-1.5) node[below,black,scale=1.pt] {$\sigma_k$} to (0,1.5) node[above,black,scale=1.pt] {$\sigma_l$};
 \node (h) at (0,1.5) [circle,draw=blue!100, fill=blue!100] [scale=0.4pt] {};
 \node (h) at (0,-1.5)  [circle,draw=blue!100, fill=blue!100] [scale=0.4pt] {};
 \draw [->,dashed,color=black] (-1.5,-1.5) node[below,scale=.8pt] {$p_m$} to (1.5,1.5) ;
       \draw [->,dashed,color=black] (1.5,-1.5) node[below,scale=.8pt] {$q_n$} to (-1.5,1.5) ;  
        		\end{tikzpicture}	
\caption{$\mathcal{W}_{p_{m}q_{n}} \big(  \sigma _{i},\sigma _{j} \big)$ (left) and $\overline{\mathcal{W}}_{p_{m}q_{n}}(\sigma_k,\sigma_l)$.}
\label{bw}
\end{figure}
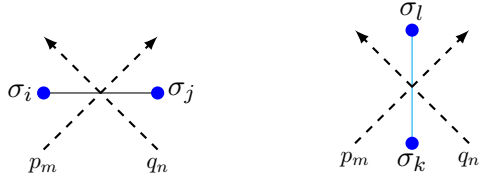

In the statistical mechanics, an Ising-type model represents a model of interacting
spins. A sketch of a simple square lattice spin model is depicted in Fig \ref{lspin}. As we can see in this figure, the spins are located at the vertices of a two-dimensional lattice, and contains $N$ sites with the spin states located on each site. The interaction between two adjacent spins $\sigma_i$ and $\sigma_j$ is described by the Boltzmann weight, $\mathcal{W}_{pq}(\sigma_i,\sigma_j)$. The parameters $p_i$ and $q_j$ are rapidities and demonstrated as dashed lines. The spins interact with their neighbours and their connections are shown by straight lines, they also interact with themselves which we call self-interactions, $\mathcal{S}(\sigma_r)$. Depending on the arrangement of the rapidity lines  with respect to the edges, Boltzmann weights are either $\mathcal{W}_{pq}(\sigma_i,\sigma_j)$ or $\overline{\mathcal{W}}_{pq}(\sigma_k,\sigma_l)$. Note that for physical systems, Boltzamann weights are real and positive. A graphical form of Boltzmann weights is given in Fig \ref{bw}. 
In the square lattice models which we consider in this paper, the spins take values, ranging, $0\leq \sigma< 2\pi$. 

The partition function of the spin lattice model is defined as the product of all Boltzmann weights and the self-interactions terms
\begin{equation}\label{Partition Function}
     \mathcal{Z}=  \int \prod_r d\sigma_r \mathcal{S}(  \sigma_r ) \prod_{i,j}\mathcal{W}_ {p_mq_n} (  \sigma _{i} ,  \sigma _{j} ) \prod_{k,l}\overline{\mathcal{W}}_ {p_mq_n}  (  \sigma _{k} ,  \sigma _{l} ), 
\end{equation} 
 where the integration is over all spin variables. Note that, for the models in which spins get discrete or discrete-continuous values, the integration has to be replaced by summation as in \cite{Baxter:1987eq}  or sum-integral as in \cite{Kels:2015bda}, respectively.

\subsection{The Star-Triangle Relation }\label{tsr}
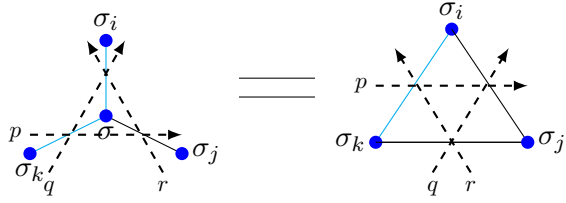
\begin{figure}[H]
\begin{tikzpicture}[scale=0.5,thick]
\draw [thin,color=black] (3.5,1) to (5.5,1) ;
  \draw [thin,color=black] (3.5,0.5) to (5.5,0.5) ;
 \draw [thin,color=cyan] (0,0)  node[below,,black,scale=1.pt] {$\sigma$} to (0,2) node[above,black,scale=1.pt] {$\sigma_{i}$};
 \node (h) at (0,2) [circle,draw=blue!100, fill=blue!100] [scale=0.4pt] {};
  \draw [thin,color=cyan] (0,0) to (-2,-1) node[below,black,scale=1.pt] {$\sigma_{k}$};
   \node (h) at (-2,-1) [circle,draw=blue!100, fill=blue!100] [scale=0.4pt] {};
     \draw [thin,color=black] (0,0) to (2,-1) node[right,scale=1.pt] {$\sigma_{j}$};
   \node (h) at (2,-1) [circle,draw=blue!100, fill=blue!100] [scale=0.4pt] {};
     \draw [->,dashed,color=black] (-2,-0.5)  node[left,scale=.8pt] {$p$} to (2,-0.5) ;
      \draw [->,dashed,color=black] (-1.5,-1.5) node[below,scale=.8pt] {$q$} to (0.5,2) ;
       \draw [->,dashed,color=black] (1.5,-1.5) node[below,scale=.8pt] {$r$} to (-0.5,2) ;
      \node (h) at (0,0) [circle,draw=blue!100, fill=blue!100] [scale=0.4pt] {};
     		\end{tikzpicture}	
\centering
\begin{tikzpicture}[scale=0.5,thick]
 \draw [thin,color=cyan] (-2,-1)   to (0,2) node[above,black,scale=1.pt] {$\sigma_{i}$};
 \node (h) at (0,2) [circle,draw=blue!100, fill=blue!100] [scale=0.4pt] {};
 \draw [thin,color=black] (0,2)   to (2,-1) ;
  \draw [thin,color=black] (2,-1) to (-2,-1) node[left,scale=1.pt] {$\sigma_{k}$};
   \node (h) at (-2,-1) [circle,draw=blue!100, fill=blue!100] [scale=0.4pt] {};
     \draw [thin,color=black] (-2,-1) to (2,-1) node[right,scale=1.pt] {$\sigma_{j}$};
   \node (h) at (2,-1) [circle,draw=blue!100, fill=blue!100] [scale=0.4pt] {};
     \draw [->,dashed,color=black] (-2.,0.5)  node[left,scale=.8pt] {$p$} to (2,0.5) ;
      \draw [->,dashed,color=black] (-0.5,-1.8) node[below,scale=.8pt] {$q$} to (1.5,1.5) ;
       \draw [->,dashed,color=black] (0.5,-1.8) node[below,scale=.8pt] {$r$} to (-1.5,1.5) ;
      		\end{tikzpicture}	

\caption{The star-triangle relation}
\label{st}
 \end{figure}

\linespread{0.5}
The star-triangle relation is the distinguished form of the Yang-Baxter equation. The solution to it suffices to ensure the integrabilty of a system. The solutions to the star-triangle relation representing  integrable spin models with discrete spin variables are presented in \cite{Kashiwara:1986tu,vonGehlen:1984bi,AuYang:1987zc,Baxter:1987eq}. The solutions with continuous spin variables can be found in \cite{Zamolodchikov:1980mb,Bazhanov:2007mh,Stroganov:1979et,Zamolodchikov:1979ba,Baxter:1982xp}.  The form of the star-triangle relation is given by
\begin{align} \nonumber
& \int d \sigma \mathcal{S}(\sigma)  \overline{\mathcal{W}}_{qr}(\sigma,\sigma_i)\mathcal{W}_{pr}(\sigma_j,\sigma) \overline{\mathcal{W}}_{pq}(\sigma_k,\sigma) =\\
& \qquad \qquad \qquad \qquad \qquad  \mathcal{R}(p,q,r)\mathcal{W}_{pq}(\sigma_i,\sigma_j) \overline{\mathcal{W}}_{pr}(\sigma_k,\sigma_i)\mathcal{W}_{qr}(\sigma_k,\sigma_j).
\end{align}\label{Star-tr}
A graphical form of (\ref{Star-tr}) is given in Fig \ref{st}. The star-triangle relation for the models with non-symmetric Boltzmann weights is given as 
\begin{align} \nonumber
& \int d \sigma \mathcal{S}(\sigma)  \overline{\mathcal{W}}_{qr}(\sigma_i,\sigma)\mathcal{W}_{pr}(\sigma,\sigma_j) \overline{\mathcal{W}}_{pq}(\sigma,\sigma_k) =\\
& \qquad \qquad \qquad \qquad \qquad \mathcal{R}(p,q,r)\mathcal{W}_{pq}(\sigma_j,\sigma_i) \overline{\mathcal{W}}_{pr}(\sigma_i,\sigma_k)\mathcal{W}_{qr}(\sigma_j,\sigma_k),
\end{align}
where the factor $\mathcal{R}(p,q,r)$ is independent of the spins\footnote{ In the most of the known statistical models,  $ \mathcal{R}(p,q,r)=f_{qr}f_{pq}/f_{pr},
$
where $f$'s are scalar functions of rapidity variables $p$, $q$ and $r$ \cite{Bax02rip}.}. For our new model the two forms of the star-triangle relation are the same.

 Considering the fact that in most  of the spin models, the Boltzmann weights depend on the difference between rapidity parameters, we set the two types of Boltzmann weights 
\begin{align}
    \mathcal{W}_{\alpha}(\sigma_i,\sigma_j):= \mathcal{W}_{pq}(\sigma_i,\sigma_j),\qquad\qquad
    \overline{\mathcal{W}}_{\alpha}(\sigma_i,\sigma_j):= \overline{\mathcal{W}}_{pq}(\sigma_i,\sigma_j),
\end{align}
with the spectral parameters, which is defined e.g. as $\alpha:=p-q$. \\As we mentioned before, our Boltzmann weights are symmetric, namely
\begin{equation}
\mathcal{W}_\alpha(\sigma_i,\sigma_j)=\mathcal{W}_\alpha(\sigma_j,\sigma_i), \end{equation}
and we also have 
\begin{equation}
\overline{\mathcal{W}}_ \alpha (\sigma_i,\sigma_j)=\mathcal{W}_{ \eta - \alpha }(\sigma_i,\sigma_j).
\end{equation}
Hence, the star-triangle relation reads
\begin{align}\nonumber\label{str}
     \int d \sigma \mathcal{S}( \sigma ) \mathcal{W}_{\eta-\gamma}(\sigma,\sigma_i)& \mathcal{W}_{\eta - \beta}( \sigma _{j}, \sigma) \mathcal{W}_{ \eta - \alpha}( \sigma_{k} , \sigma)=\\
    & \qquad \mathcal{R}(\alpha,\beta,\gamma)\mathcal{W}_{ \alpha }( \sigma _{i}, \sigma_{j})\mathcal{W}_{ \beta}( \sigma _{k}, \sigma_{i})\mathcal{W}_{ \gamma}( \sigma _{k}, \sigma_{j}),
\end{align}
where $\eta=\alpha+\beta+\gamma$ is the relation between spectral parameters.\\Considering the new Boltzmann weights, the partition function becomes
\begin{equation}\label{P-F}
      \mathcal{Z}=\int \prod_r d\sigma_r   \mathcal{S}(  \sigma _{r} ) \prod_{i,j}  \mathcal{W}_ {\alpha} (  \sigma _{i} ,  \sigma _{j} ) \prod_{k,l}  \mathcal{W}_ {\eta-\alpha }  (  \sigma _{k} ,  \sigma _{l} ).  
\end{equation}
 
Note that, it is possible to evaluate the partition function in thermodynamic limit, in which the number
of lattice sites goes to infinity and the bulk free energy  vanishes
\begin{equation}
    \lim_{N\rightarrow\infty} \frac{1}{N}\log  \mathcal{Z}=0.
\end{equation}
If the partition function can be evaluated without any approximation in the thermodynamic limit, it is an exactly solvable model. The star-triangle relation suffices for the integrability of our system. We can evaluate the partition function exactly for those cases in which the Boltzmann weights satisfy the star-triangle relation.

\section{Solutions to the Yang-Baxter Equation} \label{sol} \linespread{0.5}

The Yang–Baxter equation or as a special form, the star-triangle relation on the statistical mechanical side can be translated to the invariance of the supersymmetric indices on the gauge theory side under the Seiberg-type duality.
In this section, we show how the identification of the two sides, results in the new solution of the star-triangle relation, and leads to a new integrable Ising-type system. We also present the high temperature limit of this model. Based on the star-triangle relation, we demonstrate the solutions to the IRF and the vertex type YBE.

\subsection{Ising-type Model}\label{SL-Sbrg}
\linespread{0.5}

The star-triangle relation of the two-dimensional spin model with continuous spin variables which we discussed in \ref{tsr} reads
\begin{align}\nonumber\label{str}
     \int d \sigma \mathcal{S}( \sigma ) \mathcal{W}_{\eta-\gamma}(\sigma,\sigma_i)& \mathcal{W}_{\eta - \beta}( \sigma _{j}, \sigma) \mathcal{W}_{ \eta - \alpha}( \sigma_{k} , \sigma)=\\
    & \qquad \mathcal{R}(\alpha,\beta,\gamma)\mathcal{W}_{ \alpha }( \sigma _{i}, \sigma_{j})\mathcal{W}_{ \beta}( \sigma _{k}, \sigma_{i})\mathcal{W}_{ \gamma}( \sigma _{k}, \sigma_{j}).
\end{align}
This is the special form of the YBE and plays the role of statistical side. In order  to get  an integrable Ising-type model  using  the gauge/YBE correspondence, we identify the supersymmetric indices in Seiberg-type duality with the star-triangle relation. The Seiberg-type duality consists of two parts. \\
The electric part, in which the theory has the gauge group $SU(2)$ and the flavor $SU(6)$. The chiral multiplets transform under fundamental representation of the gauge and flavor group. The vector multiplets transform in the adjoint representation of the gauge group, and 
the magnetic part, in which there is no gauge symmetry and there are 15 chiral multiplets, and they transform under totally antisymmetric tensor representation of the flavor group.

The supersymmetric indices which are computed from the vector and chiral multiplets are taken from \cite{Gadde:2013ftv}. Therefore, the identification of the  2d $\mathcal{N}$=(2,2) supersymmetric indices of the Seiberg-type dual theories, and using formulas (\ref{Chiral}) and (\ref{Vec}), leads to the equality

\begin{align}\label{Seiberg}
    \frac{1}{2}\Big(\frac{(q;q)^2_{\infty}}{\theta(y;q)}\Big) \int \frac{dz}{2\pi i z} \frac{\prod_{i=1}^6\Delta(a_iz^{\pm1};q,y)}{\Delta(z^{\pm2};q,y)}
     =\prod_{1\leq i<j\leq 6}\Delta(a_ia_j;q,y),
\end{align}
where $\theta$ is the Jacobi theta function (\ref{theta}), and $(q;q)_\infty$ is $q$-Pochmmher (\ref{q-Poch}). As for definition
\begin{equation}
 \theta(az^{\pm};q):= \theta(az;q) \theta(az^{-1};q),
\end{equation}
 and we also have 
\begin{align} \label{del}
  \Delta(a;q,y):=\frac{\theta(ay;q)}{\theta(a;q)}.\quad\,\,\,\,\,
\end{align}
The balancing condition is given by
\begin{equation}
    \prod_{i=1}^6 a_i= \frac{q}{y}.\quad\,\,\,\,\,\,\,\,
\end{equation}
In order to show the solution to the star-triangle relation, let us now set the  flavor fugacities as follows
\begin{align}
    a_1= e^{-\alpha+i\sigma_i};\; a_2= e^{-\alpha-i\sigma_i};\;a_3= e^{-\beta+i\sigma_j};\;   a_4= e^{-\beta-i\sigma_j}\\ \nonumber
     a_5= e^{-\gamma+i\sigma_k};\; a_6 = e^{-\gamma-i\sigma_k};\; \frac{q}{y}= e^{-2\eta};\;      z= e^{i\sigma}  .
\end{align}
Along with some properties of Jocobi theta function (Appendix \ref{appa}) which are useful to obtain the new solution, we take  the main step and identify  the star-triangle relation in (\ref{str}) with the equality  given in (\ref{Seiberg}). Therefore, we obtain  the Boltzmann weights, the self-interactions and the R-factors of the new integrable spin model as follows
\begin{equation}\label{B-W}
     \mathcal{W}_{\alpha}(\sigma_i,\sigma_j)=\frac{\theta( e^{-\alpha-\eta\mp i(\sigma_i\pm \sigma_j)};q)}{\theta(e^{\alpha-\eta\pm i(\sigma_i\pm \sigma_j)};q)},\qquad\qquad\qquad\qquad\,\,\,\,\,\,
\end{equation}

\begin{equation}\label{S}
     \mathcal{S}(\sigma)=\frac{1}{4\pi}\Big(\frac{(q;q)^2_{\infty}}{\theta(y;q)}\Big)  \frac{\theta(e^{\pm 2i\sigma};q)}{\theta(e^{-2\eta\pm 2i\sigma};q)}, \qquad\quad\,\,\,
\end{equation}

\begin{equation}\label{R}
    \mathcal{ R}(\alpha,\beta,\gamma)= \frac{\theta(e^{-2\alpha-2\gamma};q)}{\theta(e^{-2\beta};q)} \frac{\theta(e^{-2\beta-2\gamma};q)}{\theta(e^{-2\alpha};q)} \frac{\theta(e^{-2\beta-2\alpha};q)}{\theta(e^{-2\gamma};q)}.
\end{equation}
These results correspond to a new Ising-type model. Note that, considering the same duality, there has been found other solutions to the star-triangle relation with continuous spin in \cite{Spiridonov:2010em}, and discrete-continuous spin variables in \cite{Kels:2015bda,Gahramanov:2015cva}.\\ It can be shown that
the normalization conditions read
\begin{align} \label{nc}\nonumber
      \mathcal{W}_{\alpha\rightarrow 0}(\sigma_i,\sigma_j)=1,  \qquad \qquad \qquad  \qquad\qquad\qquad\qquad \,\,\,\,\,
     \\
      \mathcal{W}_{\eta-\alpha}|_{\alpha\rightarrow 0}(\sigma_i,\sigma_j)= \Big(\frac{\theta(y;q)}{(q;q)^2_{\infty}}\Big) \frac{(\delta(\sigma_i-\sigma_j)+\delta(\sigma_i+\sigma_j))}{\mathcal{S}(\sigma_i)}.
\end{align}

Using these normalization conditions, and the star-triangle relation (\ref{str}), it can be shown that the  weights  satisfy the  \emph{inversion relations} 
\begin{align}\label{I-R}
    \mathcal{S}(\sigma_i)\int d \sigma \overline{\mathcal{W}}_{pq}(\sigma_i,\sigma)\mathcal{S}(\sigma) \overline{\mathcal{W}}_{qp}(\sigma,\sigma_j)=f_{pq}f_{qp} \delta_{\sigma_i,\sigma_j}, \qquad\quad\\\nonumber
    \mathcal{W}_{pq}(\sigma_i,\sigma_j)\mathcal{W}_{qp}(\sigma_i,\sigma_j)=1, \qquad \forall \sigma_i,\sigma_j \in R. 
\end{align} 
Note that, the star-triangle relation (\ref{Star-tr}) can be expressed without R-factor $\mathcal{R}(p,q,r)$, if the Boltzmann weights in (\ref{B-W}), are normalized by $k(\alpha)$ as follows
\begin{equation}\label{NB-W}
     \mathcal{ W}_{\alpha}(\sigma_i,\sigma_j)=\frac{1}{k(\alpha)}\frac{\theta( e^{-\alpha-\eta\mp i(\sigma_i\pm \sigma_j)};q)}{\theta(e^{\alpha-\eta\pm i(\sigma_i\pm \sigma_j)};q)},\qquad\qquad \quad
\end{equation}
where 
\begin{align}\label{K}
     k(\alpha)= \frac{\theta(e^{-2\alpha-\eta};q)}{\theta(e^{2\alpha-\eta};q)},\;\;\ k(\alpha) k(-\alpha)=1.
\end{align}
In this case, the partition function per site reads 
 \begin{equation}
     \lim _{N\rightarrow\infty} \mathcal{Z}^{\frac{1}{N}}=1.
      \end{equation}

\subsubsection{The High Temperature Limit}\label{ht}
In this part, we express the high temperature limit of our new model  in terms of hyperbolic sine function, which we get it using the Seiberg-type duality (\ref{Seiberg}). This limit  corresponds to the reduction of 2d index to 1d or  compactification of index on torus to the circle \cite{Yamazaki:2015voa}. The detailed discussion of 1d index can be found in \cite{Hori:2014tda, Cordova:2014oxa}.\\To do so, let us consider the following  redefinition of variables in (\ref{del}),
\begin{align}
  a=e^{2\pi i \delta};\; q=e^{2\pi i \tau};\; y=e^{2\pi i \xi},
\end{align}
and set 
\begin{align}
  \tau=i\beta;\; \delta=\beta v;\; \xi= \beta t.
\end{align}
Hence, we have
\begin{align}\label{Deltanew}
 \Delta(a;q,y) = \prod_{n=0}^{\infty} \frac{(1-e^{-2\pi\beta(-i(v+t)+n)})(1-e^{-2\pi\beta(i(v+t)+n+1)})}{(1-e^{-2\pi\beta(-iv+n)})(1-e^{-2\pi\beta(iv+n+1)})},\qquad\qquad\quad
\end{align}
then taking the  $\beta\rightarrow0$ limit, it leads to
\begin{align}
\lim_{\beta \to 0} \Delta(a;q,y)=  \prod_{n=0}^{\infty} \frac{(-i(v+t)+n)}{(-iv+n)} \frac{(i(v+t)+n+1)}{(iv+n+1)}= \\ \nonumber
\frac{(v+t)}{ v} \prod_{n=1}^{\infty} \frac{1+\frac{(v+t)^2}{n^2}}{1+\frac{v^2}{n^2}} =
\frac{\sinh\pi(v+t)}{\sinh\pi v},
\end{align}
where we use the infinite product representation 
\begin{align}
  \sinh x=x\prod_{n=1}^{\infty}(1+\frac{x^2}{\pi^2n^2}),
\end{align}
and similarly we have
\begin{align}
     \lim_{\beta \to 0} \frac{(q;q)^2_{\infty}}{\theta(y;q)}=\frac{\pi i}{\sinh(\pi t)}.\qquad\qquad\quad
\end{align}
Therefore, the dimensional reduction of (\ref{Seiberg}) becomes
\begin{align}\label{sinhstr}\nonumber
  \frac 12 \frac{\pi i}{\sinh(\pi t)} \int dx \frac{\sinh(\pm 2\pi x) }{\sinh \pi(\pm 2x+t)}\prod_{i=1}^{6} &\frac{\sinh\pi (a_i\pm x+t) }{\sinh \pi(a_i\pm x)}=\\&\qquad\qquad\prod_{1\leq i<j\leq6}\frac{\sinh\pi(a_i+a_j+t)}{\sinh\pi(a_i+a_j)}, 
\end{align}
where 
\begin{align}
   \sinh(\pm x-t):=\sinh(x-t)\sinh(-x-t),
\end{align}
and the balancing condition of the integral identity (\ref{sinhstr}) reads
\begin{align}
    \sum_{i=1}^6 a_i=1.
\end{align}
At this step, if we redefine the parameters in (\ref{sinhstr}) as follows 
\begin{align}
  a_1=-\alpha+i\sigma_i;\;\; a_2=-\alpha-i\sigma_i;\;\; a_3=-\beta+i\sigma_j;\\ \nonumber
     a_4=-\beta-i\sigma_j;\;\; a_5=-\gamma+i\sigma_k;\;\; a_6=-\gamma-i\sigma_k; \;\; x=i\sigma,
\end{align}
we get a new solution to the star-triangle relation with continuous spin variables in terms of hyperbolic sine function. The new Boltzmann weights are
\begin{equation}
   \mathcal{W}_\alpha(\sigma_i,\sigma_j)=\frac{\sinh\pi((-\eta+\alpha)\pm i(\sigma_i\pm\sigma_j)+t)}{\sinh\pi((-\eta+\alpha)\pm i(\sigma_i\pm\sigma_j))},\qquad\qquad\qquad\,\,\,\,\,
\end{equation}
\begin{equation}
    S(\sigma)=\frac 12 \frac{-\pi}{\sinh\pi t}\frac{\sinh\pi(\pm 2 i\sigma) }{\sinh \pi(\pm 2i\sigma+t)}, \qquad \qquad\qquad\qquad
    \end{equation}
\begin{equation}
  R(\alpha,\beta,\gamma) =\frac{\sinh\pi(-2\alpha+t)}{\sinh\pi(-2\alpha)}\frac{\sinh\pi(-2\beta+t)}{\sinh\pi(-2\beta)}\frac{\sinh\pi(-2 \gamma+t)}{\sinh\pi(-2\gamma)}.\,\,\,\,
\end{equation}

\subsection{Interaction-Round-a-Face Model}
\linespread{0.5}
In this part, we firstly describe the IRF model, and then present the solution we obtain, using the star-triangle relation in (\ref{str}).

\begin{figure}[H] 
\centering
\begin{tikzpicture}[scale=0.5,thick]
\node (h) at (0,2) [circle,draw=blue!100, fill=blue!100] [scale=0.5pt] {};
\node (h) at (0,-2) [circle,draw=blue!100, fill=blue!100] [scale=0.5pt] {};
\node (h) at (-2,0) [circle,draw=blue!100, fill=blue!100] [scale=0.5pt] {};
\node (h) at (2,0) [circle,draw=blue!100, fill=blue!100] [scale=0.5pt] {};
\draw [thin,color=black] (0,2) to (-2,0)  node[above,scale=1.pt] {${a}$} ;
\draw [thin,color=black] (-2,0) to (0,-2)  node[above,scale=1.pt] {${d}$} ;
\draw [thin,color=black] (0,-2) to (2,0)  node[above,scale=1.pt] {${c}$} ;
\draw [thin,color=black] (2,0) to (0,2)  node[above,scale=1.pt] {${b}$} ;
\draw [->,dashed,color=black] (2,-2) node[below,scale=1.pt] {$(t_k,t_l)$} to (-2,2)  ;
\draw [->,dashed,color=black] (-2,-2) node[below,scale=1.pt] {$(t_i,t_j)$} to (2,2)  ;
\end{tikzpicture}
\caption{Face weight}
\label{fw}
 \end{figure}
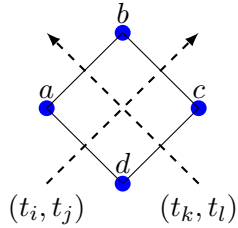

In the IRF system, the spins or spin states take integer values, which designated by variables $\{a, b, . . .\}$. The four spins surrounding a site of the lattice interact via a
 \textit{face weight}, $\mathbb{R}_{{{(t_i,t_j)(t_k,t_l)}}}\left(\begin{array}{cc}
{a} & {b} \\
{c} & {d}\end{array}\right)$.  The face weight is the product of local Boltzmann weights, associated
to $2\times2$-blocks of the adjacent entries in the array. The arrangement of the spins ${a}$, ${b}$, ${c}$, ${d}$ in a face weight are depicted in Fig \ref{fw}, and their location demonstrated by the solid circles. If we consider the interaction between all face weights, in an arbitrary spin arrangements, the model is called the interaction-round-a-face.  The system can be seen as a model in which the  states are
rectangular arrays of fixed size with real and integer entries. The real entries are regarded as 
heights of the surface over the rectangle.
  
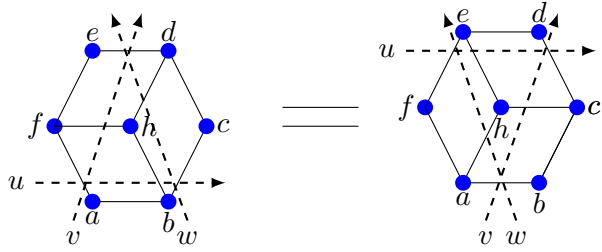
\begin{figure}[H]
 \begin{tikzpicture}[scale=0.5,thick]
\draw [thin,color=black] (4,0) to (6,0) ;
  \draw [thin,color=black] (4,0.5) to (6,0.5) ;
 \draw [thin,color=black] (0,0)  node[right,scale=1.pt] {${h}$} to (1,2) node[above,scale=1.pt] {$d$};
 \node (h) at (-2,0) [circle,draw=blue!100, fill=blue!100] [scale=0.5pt] {};
  \draw [thin,color=black] (0,0) to (-2,0) node[left,scale=1.pt] {${f}$};
  \draw [thin,color=black] (0,0) to (1,-2) node[below,scale=1.pt] {${b}$};
  \draw [thin,color=black] (-2,0) to (-1,-2) node[below,scale=1.pt] {${a}$};
  \draw [thin,color=black] (-1,-2) to (1,-2) node[below,scale=1.pt] {};
  \draw [thin,color=black] (-2,0) to (-1,2) node[above,scale=1.pt] {${e}$};
  \draw [thin,color=black] (-1,2) to (1,2) node[below,scale=1.pt] {};
  \draw [thin,color=black] (1,2) to (2,0) node[right,scale=1.pt] {${c}$};
  \draw [thin,color=black] (1,-2) to (2,0) node[below,scale=1.pt]{} ;
   \node (h) at (1,-2) [circle,draw=blue!100, fill=blue!100] [scale=0.5pt] {};
   \node (h) at (-1,-2) [circle,draw=blue!100, fill=blue!100] [scale=0.5pt] {};
   \node (h) at (1,2) [circle,draw=blue!100, fill=blue!100] [scale=0.5pt] {};
   \node (h) at (-1,2) [circle,draw=blue!100, fill=blue!100] [scale=0.5pt] {};
   \node (h) at (2,0) [circle,draw=blue!100, fill=blue!100] [scale=0.5pt] {};
              \node (h) at (0,0) [circle,draw=blue!100, fill=blue!100] [scale=0.5pt] {};
                  
  \draw [->,dashed,color=black] (-2.5,-1.5)  node[left,scale=1.pt] {$u$} to (2.5,-1.5) ;
      \draw [->,dashed,color=black] (-1.5,-2.5) node[below,scale=1.pt] {$v$} to (0.3,3) ;
       \draw [->,dashed,color=black] (1.5,-2.5) node[below,scale=1.pt] {$w$} to (-0.5,3) ;
		\end{tikzpicture}	
\centering
\begin{tikzpicture}[scale=0.5,thick]
\draw [thin,color=black] (0,0)  node[below,scale=1.pt] {${h}$} to (-1,2) ;
 \node (h) at (-2,0) [circle,draw=blue!100, fill=blue!100] [scale=0.5pt] {};
  \draw [thin,color=black] (0,0) to (-1,-2) node[below,scale=1.pt] {${a}$};
  \draw [thin,color=black] (0,0) to (2,0) node[right,scale=1.pt] {${c}$};
  \draw [thin,color=black] (2,0) to (1,-2) node[below,scale=1.pt] {${b}$};
  \draw [thin,color=black] (-1,-2) to (1,-2) node[below,scale=1.pt] {};
  \draw [thin,color=black] (-2,0) to (-1,2) node[above,scale=1.pt] {${e}$};
  \draw [thin,color=black] (-1,-2) to (-2,0) node[left,scale=1.pt] {${f}$};
  \draw [thin,color=black] (-1,2) to (1,2) node[above,scale=1.pt]{$d$} ;
  \draw [thin,color=black] (1,2) to (2,0) node[right,scale=1.pt] {${c}$};
  \draw [thin,color=black] (1,-2) to (2,0) node[below,scale=1.pt]{} ;
   \node (h) at (1,-2) [circle,draw=blue!100, fill=blue!100] [scale=0.5pt] {};
   \node (h) at (-1,-2) [circle,draw=blue!100, fill=blue!100] [scale=0.5pt] {};
   \node (h) at (1,2) [circle,draw=blue!100, fill=blue!100] [scale=0.5pt] {};
   \node (h) at (-1,2) [circle,draw=blue!100, fill=blue!100] [scale=0.5pt] {};
   \node (h) at (2,0) [circle,draw=blue!100, fill=blue!100] [scale=0.5pt] {};
     
              \node (h) at (0,0) [circle,draw=blue!100, fill=blue!100] [scale=0.5pt] {};
  
  \draw [->,dashed,color=black] (-2.5,1.5)  node[left,scale=1.pt] {$u$} to (2.5,1.5) ;
      \draw [->,dashed,color=black] (0.4,-3) node[below,scale=1.pt] {$w$} to (-1.5,2.5) ;
       \draw [->,dashed,color=black] (-0.4,-3) node[below,scale=1.pt] {$v$} to (1.5,2.5)  ;
		\end{tikzpicture}	
\caption{The Yang-Baxter equation for IRF model}
\label{IRF YBE}
 \end{figure}

There are several ways to get IRF models from Ising-type models. In our case, we use the star-triangle movements, i.e. we apply three times the star-triangle relation, and once the triangle-star  relation. The detailed derivation is given in Appendix \ref{AIRF}.\\
For our new model, the IRF type YBE with continuous spin values,  gets the following form
\begin{align}\label{irf}
\begin{array}{l}
\displaystyle  \int [dh]\;
\mathbb{R}_{{{(t_4,t_1)(t_6,t_3)}}}\left(\begin{array}{cc}
{a} & {b} \\
{h} & {c}\end{array}\right)
\mathbb{R}_{{(t_6,t_3)(t_2,t_5)}}\left(\begin{array}{cc}
{c} & {d} \\
{h} & {e}\end{array}\right)
\mathbb{R}_{{(t_2,t_5)(t_4,t_1)}}\left(\begin{array}{cc}
{e} & {f} \\
{h} & {a}\end{array}\right)=\\
[5mm]
\displaystyle \;\;\;\;\;\;\;\;\;\;\;\;\;\;\; \int [dh]\;
\mathbb{R}_{{(t_6,t_3)(t_2,t_5)}}\left(\begin{array}{cc}
{b} & {h} \\
{a} & {f}\end{array}\right)
\mathbb{R}_{{(t_2,t_5)(t_4,t_1)}}\left(\begin{array}{cc}
{d} & {h} \\
{c} & {b}\end{array}\right)
\mathbb{R}_{{(t_4,t_1)(t_6,t_3)}}\left(\begin{array}{cc}
{f} & {h} \\
{e} & {d}\end{array}\right),\qquad
\end{array}\qquad\qquad\qquad\qquad\qquad
\end{align}
where $(t_i,t_j)$  are the spectral parameters. A graphical representation of IRF type YBE is given in Fig \ref{IRF YBE}. 
 The appropriate face weights of the new model read 
\begin{align}\label{sirf}
    \mathbb{R}_{(t_i,t_j)(t_k,t_l)}\left( \begin{array}{cc} {a} & {}{b}\\{f}&{}{h} \end{array}\right) = \nonumber \frac{1}{2}\Big(\frac{(q;q)^2_{\infty}}{\theta(y;q)}\Big) \Delta(e^{2t_l-2t_k+\frac{2}{3}};q,y)\Delta(e^{2t_j-2t_i+\frac{2}{3}};q,y) \times\\ \nonumber
     \int \frac{dz_i}{2\pi i z_i \Delta(z^{\pm2};q,y)} \mathcal{W}_{\frac{1}{6}+t_i-t_l}(a;z_i)  \mathcal{W}_{\frac{1}{3}+t_j-t_i}(b;z_i) \times\\ \,\,\,\,\,\,\,\,\mathcal{W}_{\frac{1}{3}+t_l-t_k}(f;z_i)\mathcal{W}_{\frac{1}{6}+t_k-t_j}(h;z_i),
       \end{align}
where $\mathcal{W}$ and $\mathcal{S}$ are given in (\ref{B-W}) and (\ref{S}), respectively.\\
The  multi-spin  generalization of the IRF model which we  construct from the star-triangle relation has been studied in \cite{Yamazaki:2015voa}.

\subsection{Vertex Model}
\begin{figure}[H]
  \centering
\begin{tikzpicture}[scale=0.5,thick]
  \draw [->, thin,color=black] (-2,-2)  node[below,scale=1.pt] {$u_1$} to (2,2) ; 
  \draw [->, thin,color=black] (2,-2) node[below,scale=1.pt] {$u_3$}    to (-2,2) ;
  \draw [->, thin,color=black] (1,-2.5)  node[left,scale=1.pt] {$u_2$} to (1,2.5) ;
  \draw [thin,color=black] (3,0) to (5,0) ;
  \draw [thin,color=black] (3,-0.5) to (5,-0.5) ;
    \end{tikzpicture}
        \begin{tikzpicture}[scale=0.5,thick]
      \draw [->, thin,color=black] (-2,-2) node[below,scale=1.pt] {$u_1$} to (2,2)  ;
  \draw [->, thin,color=black] (2,-2) node[below,scale=1.pt] {$u_3$} to (-2,2) ;
  \draw [->, thin,color=black]  (-1,-2.5) node[right,scale=1.pt] {$u_2$}  to (-1,2.5) ; 
    \end{tikzpicture}
  \caption{Vertex type YBE}
 \label{mylabel}
 \end{figure}
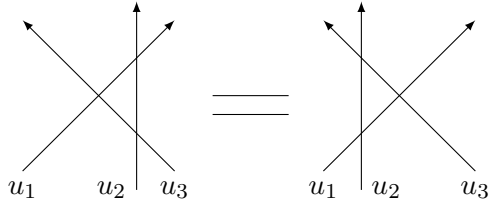

\linespread{0.5}
The vertex type YBE presents a natural way to make sense of triple crossings. To have an integrable model defined on a Baxter lattice\footnote{A Baxter lattice is any
lattice that can be drawn in a plane as a collection of lines  that undergoes only pairwise intersections.}, the vertex type YBE has to be  satisfied.  If we consider the tensor product of three infinite-dimensional
spaces $V_1\otimes V_2\otimes V_3$, then associate with each space $V_i$ the spectral variable $u_i$, and the spin variable $g_i$, respectively, the vertex type YBE gets 
\begin{align} \nonumber\label{Vertex}
    \mathbb{R}_{12}(u_1,g_1&|u_2,g_2) \mathbb{R} _{13}(u_1,g_1|u_3,g_3) \mathbb{R} _{23}(u_2,g_2|u_3,g_3)=\\\qquad & \mathbb{R} _{23}(u_2,g_2|u_3,g_3)  \mathbb{R} _{13}(u_1,g_1|u_3,g_3)  \mathbb{R}_{12}(u_1,g_1|u_2,g_2),
\end{align}
where $\mathbb{R}_{ij}(u_i,g_i|u_j,g_j)$is called $\mathbb{R}$-operator, and acts in a non-trivial way in the subspace $V_i\otimes V_j$ with the unity
operator acting in its complement. The $\mathbb{R}$-operator depends on the difference of the spectral parameters
\begin{equation}
\mathbb{R}_{ij}(u_i,g_i|u_j,g_j)=\mathbb{R}_{ij}(u_i-u_j),\qquad\qquad\,\,
\end{equation}
and we can write the vertex type YBE as follows
 \begin{align}\label{YBE}
     \mathbb{R}_{12}(u_1-u_2) \mathbb{R}_{13}(u_1-u_3) \mathbb{R}_{23}(u_2-u_3)= \mathbb{R}_{23}(u_2-u_3)\mathbb{R}_{13}(u_1-u_3)\mathbb{R}_{12}(u_1-u_2),  
 \end{align}
or equivalently
\begin{align}
     \mathbb{R}_{12}(u) \mathbb{R}_{13}(v) \mathbb{R}_{23}(v-u)= \mathbb{R}_{23}(v-u)\mathbb{R}_{13}(v)\mathbb{R}_{12}(u), 
 \end{align}
 where we set $u:=u_1-u_2$ and $v:=u_1-u_3$. In this case,
each line is associated to a spectral
parameter. The degrees of freedom live on the edges of the lattice, and the interactions take place at the vertices.
 A graphical form of vertex type YBE is given in Fig \ref{mylabel}. 
 
In order to express the vertex type YBE in terms of R-matrices, $R_{ij}$, we set
\begin{equation}
    \mathbb{R}_{ij}(u)= \mathbb{P}_{ij}R_{ij}(u),
\end{equation}
where $\mathbb{P}_{ij}$ is used to interchange the spaces, namely $\mathbb{P}_{ij}(\mathbb{V}_i\otimes\mathbb{V}_j)=\mathbb{V}_j\otimes\mathbb{V}_i$.\\ Hence, the
YBE in (\ref{YBE}) gets the following form
\begin{align}
   \text{R}_{23}(u,v)\text{R}_{12}(u,w)\text{R}_{23}(v,w)=\text{R}_{12}(v,w)\text{R}_{23}(u,w)\text{R}_{12}(u,v).
 \end{align}
 Following the same procedure as in \cite{Derkachov:2012iv,Gahramanov:2015cva}, we obtain the solution for the vertex type YBE.\\
 Therefore, $\text{R}_{ij}$ corresponding to our vertex model, gets the following forms 
 \begin{equation}
    \text{R}_{12}(u,w)\equiv \text{R}_{12}(\textbf{t})=\mathcal{D}(t_1/t_4)\mathcal{M}(t_1/t_3)\mathcal{M}(t_2/t_4)\mathcal{D}(t_2/t_3),
 \end{equation}
  \begin{equation}
     \text{R}_{23}(u,v)\equiv\text{R}_{23}(\textbf{t})=\mathcal{D}(t_3/t_6)\mathcal{M}(t_3/t_5)\mathcal{M}(t_4/t_6)\mathcal{D}(t_4/t_5),
 \end{equation}
 where
 $\mathcal{D}$ and $\mathcal{M}$ read 
  \begin{align}\label{D}
     \mathcal{D}(t;x,z)=\Delta(q^{\frac{1}{2}}t^{-1}x^{\pm}z;q,t^{-1}),\qquad\qquad\qquad\qquad\quad\,\,\,\,
    \end{align}
\begin{equation}\label{M}
    \mathcal{M}(t)f(z) =  \frac{1}{2}\Big(\frac{(q;q)_{\infty}}{\theta(t;q)}\Big) \int \frac{dx\,\,\Delta(qt^{-2}z^{\pm}x;q,q^{-\frac{1}{2}}t)}{2\pi i x\Delta(x^{\pm2};q,t)}f(x),
\end{equation}
where $f(z)$ is the holomorphic test function, and we have
 \begin{align}
     t_{1,2}=e^{\pm\pi iu};\ \quad t_{3,4}=e^{\pm\pi iv};\ \quad t_{5,6}=e^{\pm\pi iw}.
 \end{align}
   
The relations (\ref{D}) and (\ref{M}) are used to get the \emph{Bailey lemma}. Note that, via the Bailey lemma we can obtain the vertex type YBE and the star-triangle relation as in \cite{Derkachov:2012iv,Gahramanov:2015cva}.\\
Here, the two functions, $\alpha(x;t)$ and $\beta(z;t)$ with complex variables $z$ and $x$, make a Bailey pair with respect to the parameter $t$ 
\begin{equation}\label{1}
    \beta(x;t)=\mathcal{M}(t)\alpha(z;t).\qquad\qquad\qquad\qquad\quad\,\,
\end{equation}
The following two functions make also Bailey pair with respect to the parameter $st$
\begin{equation}\label{2}
    \alpha'(x;st)=\mathcal{D}(s;y;x) \alpha(x;t),\qquad\qquad\qquad\quad\,\,\,\,
\end{equation}
\begin{equation}\label{3}
    \beta'(x;st)=\mathcal{D}(t^{-1};y;x)\mathcal{M}(s)\mathcal{D}(st;y;z)\beta(z;t).
\end{equation}
Considering (\ref{1}), (\ref{2}) in (\ref{3}), we  get the following identity 
\begin{equation}
     \mathcal{D}(t;y;x)\mathcal{M}(st)\mathcal{D}(s;y;z)=\mathcal{M}(s)\mathcal{D}(st;y;z)\mathcal{M}(t),
\end{equation}
which is actually the star-triangle relation in (\ref{Seiberg}).

\section{Future Directions} \label{fu}

To widen the scope of our investigation, there are several directions to probe, such as
\begin{itemize}
\item It would be interesting to check the existence of phase transition for the new Ising-type model, proposed in this paper.

\item The inversion relation in (\ref{I-R}) can also be obtained by breaking the chiral symmetry as it is discussed in \cite{Spirdonov-2017}.

\item Taking the semi-classical limit of our  model, we can obtain the two-point and one-point Lagrangians similar to the  models in \cite{Bazhanov:2016ajm,Bazhanov:2010kz}.

\item In order to understand the structure in the context of brane tiling \cite{Yagi:2016oum}, it would be interesting to find $L$-operator and finite-dimensional $R$-matrices for this model as in \cite{Chicherin:2015mfv}.

\item It is compelling  to find the relation between this model and  knot invariant via the solution of the star-triangle relation \cite{article}.

\end{itemize}
These directions seem to merit consideration, and we hope to explore them in the future.

\acknowledgments
We are indebted to Ilmar Gahramanov for introducing the problem, discussions and comments, and would like to thank Arash Arabi Ardehali, Bruce Bain, Ozgur Kelekci, Emine Seyma Kutluk and Giuseppe Mussardo for comments or discussions. We wish to acknowledge the organizers of Supersymmetry and Applications Winter School, YEFIST and 24th IFG at Feza G\"{u}rsey Center, YTU and IYTE (Jan, May and Jun 2017), respectively, where we present the contents of this paper. Sh.J extends the appreciation to the hospitality of Max Planck Institute for Gravitational Physics (Albert Einstein Institute) and the Nesin Mathematical Village, where part of the project was conducted.

\appendix

\section{ The Jacobi Theta Function \label{appa} }  

In this part, we present some useful properties of the Jacobi theta function following \cite{Rosengren:2016qtr, Mod-Analyses}. \\
The theta function in terms of the q-Pochhammer is
\begin{equation}\label{theta;poch}
    \theta(z;q)=(z;q)_{\infty}(qz^{-1};q)_{\infty}.\qquad\qquad\,\,\,\,\,\,\,\,\,
\end{equation}
We can show that
\begin{equation}\label{eq:ge}
    \theta(zq;q)= -\frac{1}{z} \theta(z;q),\qquad\qquad\qquad\qquad\,\,\,\,\,\,\,\,
\end{equation}
and as a general form, it can be shown
\begin{equation}
    \theta(q^kz;q)=(-1)^k q^{-\binom{k}{2}}z^{-k}\theta(z;q), \qquad k\in Z.
\end{equation}
In addition, we have
\begin{equation} \label{eq:ge1}
    \theta(z^{-1};q)= -\frac{1}{z} \theta(z;q).\qquad\qquad\qquad\qquad\,\,\,\,\,\,\,\,
\end{equation}
Considering (\ref{eq:ge}) and (\ref{eq:ge1}), we get 

\begin{equation} \label{eq:ge2}
    \theta(z^{-1};q)=  \theta(zq;q).\qquad\qquad\qquad\qquad\,\,\,\,\,\,\,\,
\end{equation}
The following practical identities are used to prove the  (\ref{SL-Sbrg}), and they are given by
\begin{align}\label{pDelta}
    \Delta(zq;q,y)=\frac{1}{y}\Delta(z;q,y);\,\,\,\,\,\,\qquad\qquad\qquad\qquad\\ \nonumber
    \Delta(z^{-1};q,y)=\Delta^{-1}(\frac{zq}{y};q,y)\qquad\qquad\qquad\qquad.
\end{align}

\section{ Proof of the 2d Seiberg-type Duality }  \label{proof}
In this Appendix, a mathematical proof of the equality (\ref{Seiberg}) is given. For more details see \cite{Rosengren:2016qtr,Spirdonov:Sh}.\\ Let us define
\begin{align}
  \text{I}(a_1,...,a_6;q):=\int \frac{dz}{2\pi i z} \frac{\prod_{i=1}^6\Delta(a_iz^{\pm1};q,y)}{\Delta(z^{\pm2};q,y)},\qquad \qquad
     \end{align}
     and
\begin{align}
 \qquad \text{F}(a_1):=\frac{\text{I}(a_1,...,a_6;q)}{\prod_{1\leq i<j\leq 6}\Delta(a_ia_j;q,y)}.\qquad\qquad\qquad
\end{align}
By performing the same arguments in \cite{Spirdonov:Sh}, and the first property in (\ref{pDelta}), it can be seen that
\begin{align}
  \text{F}(qa_1)=\text{F}(a_1),
\end{align}
which is required for the \emph{difference equation}.
Then, we define 
\begin{align}\label{f}
  f(a_1;z):=\frac{\prod_{i=1}^6\Delta(a_iz^{\pm};q,y)}{\prod_{1\leq i<j\leq 6}\Delta(a_ia_j;q,y)\Delta(z^{\pm2};q,y)},
\end{align}
where analytic continuation of $f(a_1;z)$  gets the following integral form
\begin{align}
  \text{A}:= \int_{C} \frac{dz}{2\pi i z} f(a_1;z)=\underset{z=a_1}{\operatorname{Res}}\frac{f(a_1;z)}{z}- \underset{z=a_1^{-1}}{\operatorname{Res}} \frac{f(a_1;z)}{z}+ \int_{C'} \frac{dz}{2\pi i z} f(a_1;z).
\end{align}
$C'$ is a modification which is  running outside $z=a_1$ and inside $z=a_1^{-1}$. In the limit $ a_1\rightarrow a_2^{-1}$, $f(a_1;z)$ vanishes, so the last integral goes to zero, and we get
\begin{equation}\label{A}
\text{A} = 2 \underset{z=a_1}{\operatorname{Res}}\frac{f(a_1;z)}{z}.
\end{equation}
A simple residue calculus leads to
\begin{align}
  \underset{z=a_1}{\operatorname{Res}} \ \Delta \Big(\frac{a_1}{z};q,y \Big)= \lim_{z \to a_1} \frac{(z-a_1)\theta(\frac{a_1}{z}y;q)}{\theta(\frac{a_1}{z};q)}=\frac{a_1\theta(y;q)}{(q;q)^2_{\infty}},\qquad\qquad\qquad
\end{align}
and therefore we have
\begin{align}
  \underset{z=a_1}{\operatorname{Res}}\frac{f(a_1;z)}{z}= \frac{\theta(y;q)}{(q;q)^2_{\infty}} \frac{\prod_{i=1}^6\Delta(a_i/a_1;q,y)}{\prod_{1\leq i<j\leq 6}\Delta(a_ia_j;q,y)\Delta(a_1^{-2};q,y)}.
\end{align}
In order to evaluate (\ref{A}), the following limit is considered
\begin{align}
  \lim_{a_1 \to a_2^{-1}} \underset{z=a_1}{\operatorname{Res}}\frac{f(a_1;z)}{z}= \frac{\theta(y;q)}{(q;q)^2_{\infty}},
\end{align}
where we use the   $\prod_{i=3}^6 a_i=q/y$, and the second relation in (\ref{pDelta})
\begin{equation}
 \text{A} = \frac{2\theta(y;q)}{(q;q)^2_{\infty}},
\end{equation}\\
thus we get the equality (\ref{Seiberg}).

\section{IRF } \label{AIRF} 
In this part, using the identification of  Seiberg-type duality in (\ref{Seiberg}) with the star-triangle relation, we demonstrate the solution to the IRF type YBE.\\Let us consider the following star-triangle relation
\begin{multline} \label{s-t1}
   \qquad \qquad  \qquad\mathcal{A}\int  D_z   \mathcal{W}_{\frac{1}{6}+t-s}(a;z)  \mathcal{W}_{\frac{1}{6}+s-r}(b;z)  \mathcal{W}_{\frac{1}{6}+r-t}(c;z)
     = \\ \,\,\mathcal{K}(-2t+2s+\frac{2}{3})\mathcal{K}(-2s+2r+\frac{2}{3})\mathcal{K}(-2r+2t+\frac{2}{3}) \times\qquad\qquad\qquad\\
      \mathcal{W}_{\frac{1}{3}+t-r}(a;b) \mathcal{W}_{\frac{1}{3}+r-s}(c;a) \mathcal{W}_{\frac{1}{3}+s-t}(b;c),\\
\end{multline}
where  
 \begin{align}
     \mathcal{K}(x)=\Delta(e^{x};q,y);\quad \mathcal{A}=\frac{1}{2}\Big(\frac{(q;q)^2_{\infty}}{\theta(t;q)}\Big),
\end{align}
and $\mathcal{W}$'s are Boltzmann weights in (\ref{B-W}). Now, let us define the integration over  six variables $z_1,...,z_6$  as follows
\begin{align} \label{six}\nonumber
  \mathcal{A}^6 \int D_{z_1}... D_{z_6}  \mathcal{W}_{\frac{1}{6}+t_2-t_6}(a;z_1) \mathcal{W}_{\frac{1}{6}+t_3-t_1}(b;z_2)  \mathcal{W}_{\frac{1}{6}+t_4-t_2}(c;z_3) \times \\\nonumber  \mathcal{W}_{\frac{1}{6}+t_5-t_3}(d;z_4)W_{\frac{1}{6}+t_6-t_4}(e;z_5) \mathcal{W}_{\frac{1}{6}+t_1-t_5}(f;z_6) \times  \\\nonumber \mathcal{W}_{\frac{1}{6}+t_1-t_2}(z_2;z_1)  \mathcal{W}_{\frac{1}{6}+t_2-t_3
  }(z_3;z_2)  \mathcal{W}_{\frac{1}{6}+t_3-t_4}(z_4;z_3) \times\\ 
   \mathcal{W}_{\frac{1}{6}+t_4-t_5}(z_5;z_4) 
  \mathcal{W}_{\frac{1}{6}+t_5-t_6}(z_6;z_5)  \mathcal{W}_{\frac{1}{6}+t_6-t_1}(z_1;z_6),
 \end{align}
 then integrate over  $z_1$,  $z_3$ and $z_5$ 
 \begin{align}\label{three}\nonumber
  \mathcal{A}^3 \int D_{z_1} D_{z_3}D_{z_5}  \mathcal{W}_{\frac{1}{6}+t_2-t_6}(a;z_1)  \mathcal{W}_{\frac{1}{6}+t_4-t_2}(c;z_3) \mathcal{W}_{\frac{1}{6}+t_6-t_4}(e;z_5) \times\\\nonumber  \mathcal{A} \int D_{z_2} \mathcal{W}_{\frac{1}{6}+t_1-t_2}(z_2;z_1)  \mathcal{W}_{(\frac{1}{6}+t_2-t_3
  }(z_3;z_2) \mathcal{W}_{\frac{1}{6}+t_3-t_1}(b;z_2) \times\\\nonumber 
   \mathcal{A} \int D_{z_4} \mathcal{W}_{\frac{1}{6}+t_3-t_4}(z_4;z_3)  \mathcal{W}_{\frac{1}{6}+t_4-t_5}(z_5;z_4) \mathcal{W}_{\frac{1}{6}+t_5-t_3}(d;z_4)\times\\ 
 \mathcal{A} \int D_{z_6} \mathcal{W}_{\frac{1}{6}+t_5-t_6}(z_6;z_5) \mathcal{W}_{\frac{1}{6}+t_1-t_5}(f;z_6) \mathcal{W}_{\frac{1}{6}+t_6-t_1}(z_1;z_6),
 \end{align}
 afterwards, apply the star-triangle relation (\ref{s-t1}) for the integrals over  $z_2$,  $z_4$ and $z_6$
  \begin{multline}\label{wp}
   \,\,\mathcal{A}^3 \int D_{z_1} D_{z_3}D_{z_5}  \mathcal{W}_{\frac{1}{6}+t_2-t_6}(a;z_1) \mathcal{W}_{\frac{1}{6}+t_4-t_2}(c;z_3) \mathcal{W}_{\frac{1}{6}+t_6-t_4}(e; z_5) \times
  \\ 
  \mathcal{K}(2t_2-2t_1+\frac{2}{3})  \mathcal{K}(2t_3-2t_2+\frac{2}{3}) \mathcal{K}(2t_1-2t_3+\frac{2}{3})  
   \times \,\qquad\\
    \mathcal{W}_{\frac{1}{3}+t_3-t_2}(b; z_1) \mathcal{W}_{\frac{1}{3}+t_1-t_3}(z_1;z_3) \mathcal{W}_{\frac{1}{3}+t_2-t_1}(b;z_3) \times\,\,\,\,\\
  \mathcal{K}(2t_4-2t_3+\frac{2}{3}) \mathcal{K}(2t_5-2t_4+\frac{2}{3}) \mathcal{K}(2t_3-2t_5+\frac{2}{3}) 
  \times \qquad \\  
 \mathcal{W}_{\frac{1}{3}+t_4-t_3}(d;z_5) \mathcal{W}_{\frac{1}{3}+t_5-t_4}(d;z_3) \mathcal{W}_{\frac{1}{3}+t_3-t_5}(z_5; z_3) \times\,\,\,\,\\
  \mathcal{K}(2t_1-2t_6+\frac{2}{3}) \mathcal{K}(2t_6-2t_5+\frac{2}{3}) \mathcal{K}(2t_5-2t_1+\frac{2}{3}) \times\qquad\\
  \mathcal{W}_{\frac{1}{3}+t_1-t_6} (f;z_5) \mathcal{W}_{\frac{1}{3}+t_6-t_5}(f; z_1)
   \mathcal{W}_{\frac{1}{3}+t_5-t_1}(z_5; z_1),\\
 \end{multline}
 and apply the following triangle-star relation, which is  based on  (\ref{s-t1})
  \begin{multline}\label{wp}
  \qquad \qquad \qquad \mathcal{K}(2t_3-2t_5+\frac{2}{3})\mathcal{K}(2t_1-2t_3+\frac{2}{3})\mathcal{K}(2t_5-2t_1+\frac{2}{3})\times \\
   \mathcal{W}_{\frac{1}{3}+t_5-t_1}(z_5;z_1)  
   \mathcal{W}_{\frac{1}{3}+t_1-t_3}(z_1; z_3)   \mathcal{W}_{\frac{1}{3}+t_3-t_5}(z_5; z_3) =\qquad\\
    \mathcal{A} \int D_z \mathcal{W}_{\frac{1}{6}+t_1-t_5}(z_3; z) \mathcal{W}_{\frac{1}{6}+t_3-t_1}(z_5; z) \mathcal{W}_{\frac{1}{6}+t_5-t_3}(z_1; z),\qquad\,\\
 \end{multline}
 we get 
 \begin{multline}\label{wp1}
   \mathcal{A}^4 \int D_z D_{z_1} D_{z_3}D_{z_5}   \mathcal{K}(2t_2-2t_1+\frac{2}{3}) \mathcal{W}_{\frac{1}{6}+t_5-t_3}(z_1; z)  \mathcal{W}_{\frac{1}{3}+t_2-t_1}(b; z_3)      \times  \\ 
   \mathcal{K}(2t_3-2t_2+\frac{2}{3})  \mathcal{W}_{\frac{1}{3}+t_3-t_2}(b;z_1)  \mathcal{W}_{\frac{1}{6}+t_2-t_6}(a; z_1) \times \\ 
   \mathcal{K}(2t_4-2t_3+\frac{2}{3}) \mathcal{W}_{\frac{1}{3}+t_4-t_3}(d;z_5)   \mathcal{W}_{\frac{1}{6}+t_4-t_2}(c;z_3) \times \\ 
    \mathcal{K}(2t_5-2t_4+\frac{2}{3})  \mathcal{W}_{\frac{1}{3}+t_5-t_4}(d; z_3) \mathcal{W}_{\frac{1}{6}+t_6-t_4}(e; z_5) \times \\ 
   \mathcal{K}(2t_6-2t_5+\frac{2}{3})  \mathcal{W}_{\frac{1}{3}+t_6-t_5}(f; z_1) \mathcal{W}_{\frac{1}{6}+t_1-t_5}(z_3; z) \times \\ 
   \mathcal{K}(2t_1-2t_6+\frac{2}{3}) \mathcal{W}_{\frac{1}{3}+t_1-t_6}(f;z_5)  \mathcal{W}_{\frac{1}{6}+t_3-t_1}(z_5; z). \\
      \end{multline}
 By repeating the same procedure as in (\ref{three}) for $z_2$, $z_4$ and $z_6$ we get 
 \begin{multline}\label{wp2}
     \mathcal{A}^4 \int D_z D_{z_2} D_{z_4}D_{z_6} \mathcal{K}(2t_2-2t_4+\frac{2}{3})  \mathcal{W}_{\frac{1}{6}+t_3-t_1}(b;z_2)
     \mathcal{W}_{\frac{1}{6}+t_5-t_3}(d; z_4)  \times \qquad\qquad\qquad\,\qquad\\
\,\,\,\,\,\,\mathcal{K}(2t_4-2t_6+\frac{2}{3})     \mathcal{W}_{\frac{1}{3}+t_1-t_6}(a; z_2) \mathcal{W}_{\frac{1}{3}+t_2-t_1}(a; z_6)  \times\\
         \,\,\,\,\,\,  \mathcal{K}(2t_1-2t_3+\frac{2}{3})     \mathcal{W}_{\frac{1}{3}+t_4-t_3}(c; z_2) \mathcal{W}_{\frac{1}{3}+t_3-t_2}(c; z_4) \times\\
          \,\,\,\,\,\,\mathcal{K}(2t_5-2t_1+\frac{2}{3})  \mathcal{W}_{\frac{1}{3}+t_5-t_4}(e;z_6)    \mathcal{W}_{\frac{1}{3}+t_6-t_5}(e;z_4)  \times\\
           \,\,\,\,\,\, \mathcal{K}(2t_6-2t_2+\frac{2}{3}) \mathcal{W}_{\frac{1}{6}+t_4-t_2}(z_1; z) \mathcal{W}_{\frac{1}{6}+t_6-t_4}(z_2; z)  \times\\ \,\,\,\,\,\,\mathcal{K}(2t_3-2t_5+\frac{2}{3}) \mathcal{W}_{\frac{1}{6}+t_2-t_6}(z_4;z)\mathcal{W}_{\frac{1}{6}+t_1-t_5}(f;z_6).  \\
 \end{multline}
Using the same integral in (\ref{six}), results in the equality of (\ref{wp1}) and (\ref{wp2}). Therefore, it leads to the solution of IRF type Yang-Baxter equation.

\bibliographystyle{utphys}
\bibliography{main}
\end{document}